\documentclass[preprint]{aastex} 

\usepackage{natbib}
\usepackage{journal_abbrev}
\usepackage{lscape}
\usepackage{pdflscape}
\usepackage{amsmath}
\usepackage{subfigure}

\newcommand{\brgamma}{Br-$\gamma$~}

\begin{document}

\title{Keck Observations of the Galactic Center Source G2: Gas Cloud or Star?}
\author{K. Phifer\altaffilmark{1,6}, T. Do\altaffilmark{2,7}, L. Meyer\altaffilmark{1}, A. M. Ghez\altaffilmark{1,8}, G. Witzel\altaffilmark{1}, S. Yelda\altaffilmark{1}, A. Boehle\altaffilmark{1,8}, J. R. Lu\altaffilmark{3,9}, M. R. Morris\altaffilmark{1}, E. E. Becklin\altaffilmark{1,4}, K. Matthews\altaffilmark{5}}

\email{ghez@astro.ucla.edu}
\altaffiltext{1}{Department of Physics and Astronomy, University of California, Los Angeles}
\altaffiltext{2}{Dunlap Institute for Astronomy and Astrophysics, University of Toronto}
\altaffiltext{3}{Institute for Astronomy, University of Hawaii}
\altaffiltext{4}{NASA-Ames Research Center}
\altaffiltext{5}{Division of Physics, Mathematics, and Astronomy, California Institute of Technology}
\altaffiltext{6}{NSF Graduate Student Fellow, Grant DGE-0707424}
\altaffiltext{7}{Dunlap Fellow}
\altaffiltext{8}{Leichtman-Levine Chair in Astrophysics}
\altaffiltext{9}{Preston Fellow}
\altaffiltext{10}{NSF Astronomy and Astrophysics Postdoctoral Fellow, Grant AST-1102791}
\begin{abstract}
We present new observations and analysis of G2 -- the intriguing red
emission-line object which is quickly approaching the Galaxy's central
black hole.  The observations were obtained with the laser guide star
adaptive optics systems on the W. M. Keck I and II telescopes and
include spectroscopy ($R \sim 3600$) centered on the Hydrogen \brgamma
line as well as $K'$ (2.1 \micron) and $L'$ (3.8 \micron) imaging.
Analysis of these observations shows the \brgamma line emission has a
positional offset from the $L'$ continuum.  This offset is likely due
to background source confusion at $L'$.  We therefore present the
first orbital solution derived from \brgamma line astrometry, which
when coupled with radial velocity measurements, results in a later
time of closest approach ($2014.21 \pm 0.14$), closer periastron ($130
\ \rm{AU}$, $1900 R_{\rm{s}}$), and higher eccentricity ($0.9814 \pm
0.0060$) compared to a solution using $L'$ astrometry.  The new orbit
casts doubt on previous associations of G2 and a low surface
brightness ``tail.''  It is shown that G2 has no $K'$ counterpart down
to $K' {\sim}20 \ \rm{mag}$.  G2's $L'$ continuum and the \brgamma
line-emission is unresolved in almost all epochs; however it is
marginally extended in our highest quality \brgamma data set from 2006
and exhibits a clear velocity gradient at that time.  While the
observations altogether suggest that G2 has a gaseous component which
is tidally interacting with the central black hole, there is likely a
central star providing the self-gravity necessary to sustain the
compact nature of this object.
\end{abstract}

\subjectheadings{accretion, accretion disks -- black hole physics --
  Galaxy: center -- Galaxy: kinematics and dynamics -- infrared:
  general}

\section{Introduction}
Recently, \citet{Gillessen-etal12,Gillessen-etal13} reported the
discovery of G2, an extremely red object with spatially resolved
\brgamma emission. The object was interpreted as a ${\sim}{3}$ Earth
mass gas cloud based upon an inferred low dust temperature, observed
elongation of the \brgamma emission along the object's direction of
motion, and a claimed tail along the same orbital trajectory as
G2. This interpretation is particularly interesting because G2 is on a
highly eccentric orbit with closest approach to our Galaxy's central
black hole within the next year, potentially allowing us
to observe an unprecedented accretion event onto a supermassive black
hole and offer insight into the region surrounding the black hole
\citep[e.g.][]{Morris12,Moscibrodzka-etal12,Anninos-etal12,Saitoh-etal12,Bartos-etal13,Yusef-Zadeh-Wardle13}.

The interpretation of G2 as a gas cloud, however, is not definitive.
One challenge for the pure gas cloud scenario is that, given the
strong tidal fields in this region and G2's low self-gravity, G2 
needs to have formed quite recently ($\sim$1995, just prior to
the initial observations; \citealt{Burkert-etal12,Schartmann-etal12}).
Since such a gas cloud would be tidally disrupted during its periapse
passage in the upcoming year, the gas cloud model implies G2 will be
observed over almost the exact extent of its existence.  Therefore
several alternative scenarios invoking an underlying star have been
proposed
\citep{Miralda-Escude12,Murray-Clay-Loeb12,Scoville-Burkert13}.  In
these scenarios, the observed heated gas is posited to be
circumstellar, either intrinsic or a consequence of interaction of the
star and surrounding ambient gas.  In these stellar scenarios, G2 not
only has existed for timescales much longer than the
observed time baseline, but will also survive periapse passage.

Regardless of its nature, G2's properties and possible origin depend
critically on its orbital parameters.  These parameters have been 
estimated from observations with a short time baseline compared to
the orbital period \citep[$\sim$10 vs.$\sim$200
  years;][]{Gillessen-etal13} and in a very
crowded region, making the orbital solution susceptible to biases
\citep{Hartkopf-etal01}.  We therefore present new measurements and
analysis of G2 that minimize the effects of
source confusion on estimates of G2's orbital parameters and examine
the temporal evolution of G2's properties.

\section{Observations} \label{sec:obs}
Two types of new data were collected for this study using the
laser-guide-star adaptive optics (LGS AO) systems at the W.M. Keck
Observatory \citep{Wizinowich-etal06,vanDam-etal06,Chin-etal12}.
Spectra were obtained using the OSIRIS integral field spectrograph
\citep{Larkin-etal06} through the narrow-band Kn3 filter, which is
centered on the \brgamma hydrogen line ($2.1661 \ \mu \rm{m}$), at a
spectral resolution of $R \sim 3600$.  Imaging data were obtained in
the $K'$-band filter ($\lambda_0 = 2.124 \ \mu \rm{m}$) and the
$L'$-band filter ($\lambda_0 = 3.776 \ \mu \rm{m}$) using the Keck II
near-infrared camera (NIRC2, PI: K. Matthews).  These data were
obtained and reduced in a similar manner as in our previous
publications \citep{Lu-etal09,Yelda-etal10,LMeyer-etal12} and the
specific spectroscopic and $L'$ observations utilized in this paper
are described in table \ref{tab:obs}.

\thispagestyle{empty}
\begin{deluxetable}{lccccccccccc}
\rotate
\tabletypesize{\scriptsize}
\tablewidth{0pt}
\tablecaption{Summary of Observations and Measurements of G2\label{tab:obs}}
\tablehead{
  \colhead{UT date} & 
  \colhead{Fractional} & 
  \colhead{AO type/} & 
  \colhead{pix scale} & 
  \colhead{$N_{\rm{frames}}$} & 
  \colhead{$N_{\rm{frames}}$} & 
  \colhead{FWHM} & 
  \colhead{Orig} &
  \colhead{vlsr} & 
  \colhead{\brgamma FHWM} & 
  \colhead{G2 $\Delta$RA\tablenotemark{b}} & 
  \colhead{G2 $\Delta$Dec} \\ 
  \colhead{} & 
  \colhead{Date} & 
  \colhead{Telescope} & 
  \colhead{(mas)} & 
  \colhead{observed} & 
  \colhead{used} & 
  \colhead{(mas)} & 
  \colhead{Pub.\tablenotemark{a}} & 
  \colhead{$(\rm{km} \, \rm{s}^{-1})$} & 
  \colhead{$(\rm{km} \, \rm{s}^{-1})$} &
  \colhead{(mas)} & 
  \colhead{(mas)}
}
\startdata
\multicolumn{12}{c}{OSIRIS, Kn3} \\ \hline \hline
2006 Jun 18,30; Jul 1     & 2006.495 & Keck II LGS & 35 & 28 & 27  & 74 & 3 & $1125 \pm 9$ & $135 \pm 30$ & $222.97 \pm 5.05$ & $-105.40 \pm 2.40$ \\
2008 May 16; Jul 25       & 2008.487 & Keck II LGS & 35 & 22 & 21  & 78 & 4,0 & \tablenotemark{e} & \tablenotemark{e} & $181.82 \pm 6.90$ & $-64.97 \pm 2.51$ \\
2009 May 5,6              & 2009.344 & Keck II LGS & 35 & 24 & 19  & 79 & 0 & $1352 \pm 32$  & $171 \pm 121$   & $176.26 \pm 2.74$ & $-65.76 \pm 2.04 $ \\
2010 May 5,8              & 2010.349 & Keck II LGS & 35 & 17 & 16  & 82 & 0 & $1479 \pm 29$  & $262 \pm 165$   & $166.29 \pm 9.11$ & $-43.95 \pm 11.05 $ \\
2012 Jun 9,11; Aug 11,12  & 2012.613 & Keck I LGS & 20\tablenotemark{c} & 27 & 21\tablenotemark{d} & 68 & 0 & $2071 \pm 146$ & $706 \pm 250$ & $103.16 \pm 4.07$ & $-8.79 \pm 11.36$ \\
&&&&&&&& \\ 
\multicolumn{12}{c}{NIRC2, $L'$} \\ \hline \hline
2003 Jun 10          & 2003.440 & Keck II NGS & 10 & 12   & 12   & 86 & 0 && & $260.1 \pm 4.5$ & $-154.6 \pm 5.1$ \\
2004 Jul 26          & 2004.567 & Keck II LGS & 10 & 11   & 11   & 80 & 1 && & $262.6 \pm 4.8$ & $-140.4 \pm 4.2$ \\
2005 Jul 30          & 2005.580 & Keck II LGS & 10 & 62   & 56   & 87 & 2 && & $251.6 \pm 3.4$ & $-129.0 \pm 2.6$ \\
2006 May 21          & 2006.385 & Keck II LGS & 10 & 19   & 19   & 81 & 0 && & $226.2 \pm 1.9$ & $ -99.4 \pm 0.8$ \\
2009 Jul 22          & 2009.561 & Keck II LGS & 10 & 4    &  4   & 85 & 0 && & $175.4 \pm 4.9$ & $ -70.7 \pm 2.1$ \\
2012 Jul 20 - Jul 23 & 2012.562 & Keck II LGS & 10 & 1316 & 1314 & 93 & 0 && & $105.9 \pm 0.8$ & $ -21.1 \pm 0.6$ \\ 
\enddata
\tablenotetext{a}{References: \scriptsize(0) This Work; (1) \citet{Ghez-etal05}; (2) \citet{Hornstein-etal07}; (3) \citet{Ghez-etal08}; (4) \citet{Do-etal09}}
\tablenotetext{b}{Offset from SgrA*-radio.}
\tablenotetext{c}{A smaller square dither pattern of $0\farcs5 \times 0\farcs 5$ was used.}
\tablenotetext{d}{In 2012, a more stringent quality cut ($\rm{FWHM} < 68 \rm{mas}$) was used because G2 is closer to stars than in other epochs and because the improved LGS performance on Keck I allowed this more stringent cut.}
\tablenotetext{e}{In 2008, low SNR and changes in the local standard of rest velocity between the two observation dates prevent reliable line measurements.}
\end{deluxetable}
\section{Analysis} \label{sec:analysis}
\subsection{OSIRIS IFU Measurements}\label{sec:oskin}
Because G2 has no detectable $K$ continuum and is fainter than any
object we have previously extracted, some analysis steps differ from
the spectral extraction presented in \citet{Ghez-etal08} and
\citet{Do-etal09}.  For all epochs, we created a combined data cube
before extracting G2's spectrum rather than extracting spectra from
individual cubes.
The OH sky lines in the data are subtracted using sky frames scaled to
the strength of families of OH lines in the observed frames to account
for temporal variations.  In 2012, the \brgamma emission line from G2
is coincident with the 2.180 \micron \ OH line, so to minimize the
systematic effects associated with OH line subtraction, we scale the
sky only to this line.

\begin{figure*}[ht!]
\centering \includegraphics[angle=90,width=6in]{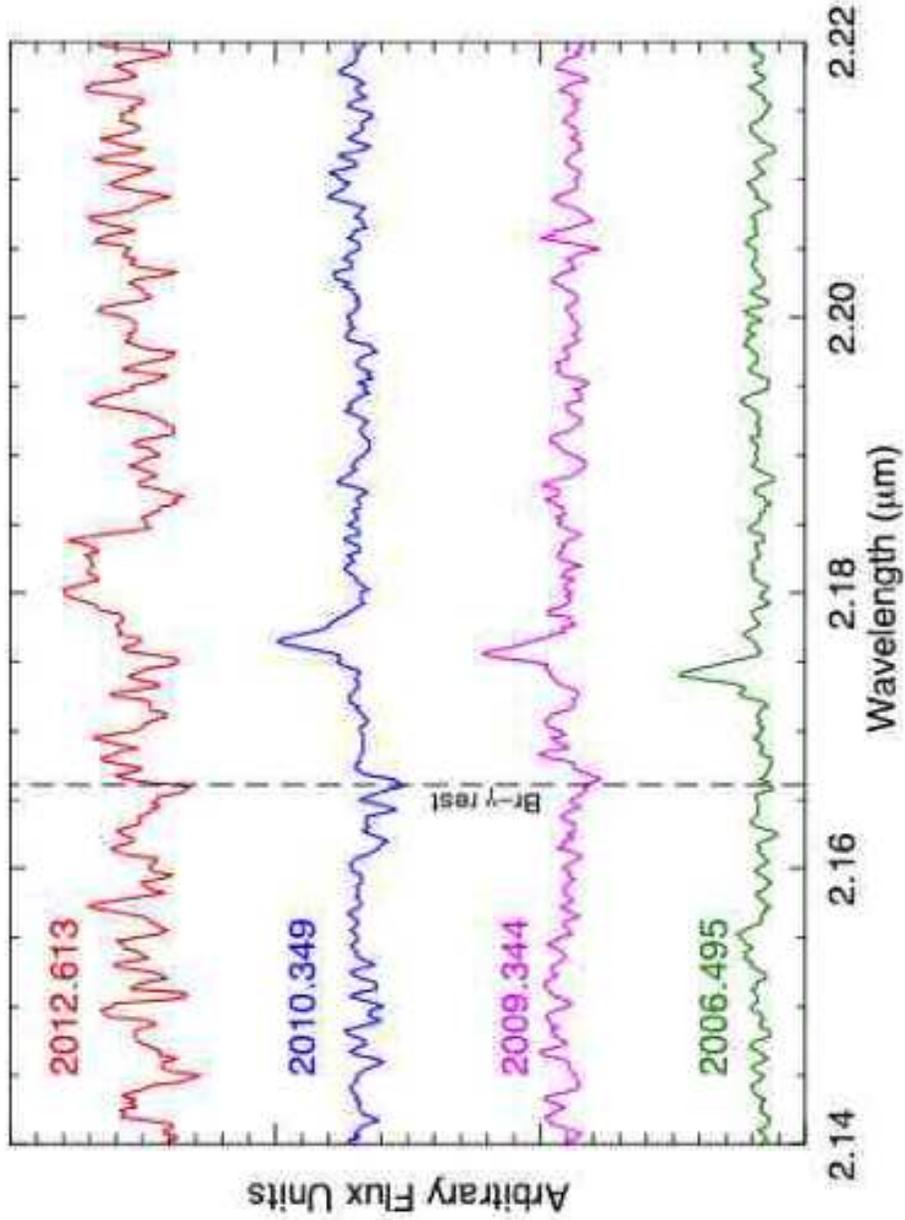}
\caption{The evolving spectra of G2.  The highly redshifted \brgamma
  emission line has a FWHM which increases with time (see table
  \ref{tab:obs}).  The \brgamma line peak has been scaled such that it
  is the same height for each spectrum.  Additionally, the spectra
  have been smoothed using a boxcar average with a width of 3 OSIRIS
  channels.}
\label{fig:G2}
\end{figure*}

An iterative process was required to estimate the position and
spectral properties of G2 in the OSIRIS data cubes (see figures
\ref{fig:G2} and \ref{fig:images}) since G2's position is needed to
place the aperture for spectral extraction, and G2's spectral
properties are needed to determine which OSIRIS channels should be
used to measure its position.  G2's position was first obtained by
visual inspection of the data cube.  Then, an initial spectrum was
extracted at this position using an aperture with a radius of 35 mas.
Emission from local ambient gas was subtracted using a region free of
stellar halos within ${\sim} 0\farcs5$. A Gaussian was fit to the
resulting emission line.  In order to refine the position, the
three-dimensional data cube was median collapsed over the wavelength
range corresponding to twice the standard deviation of the Gaussian
fit to the emission line, centered on the line peak.  Continuum
emission was subtracted by averaging the median of the 25 nearest
channels on either side of the line.  G2's position was further
refined with a two-dimensional Gaussian fit to the
continuum-subtracted image.  The iterative extraction process was
repeated again to obtain a final position and spectrum for G2.
Further iterations produced no significant changes.  Measuring G2's
position on a continuum-subtracted frame (see figure
\ref{fig:brgamma}) removes the effect of source confusion, thereby
avoiding astrometric biases.

G2's spectral properties were obtained using the final Gaussian fits
to the \brgamma emission line.  The radial velocities (RVs) for G2
were calculated from the offset of the line from the rest wavelength
($\lambda_{Br\gamma}=2.1661 \ \mu m$) and corrected to the local
standard of rest.  The reported FWHM measurements were corrected for
instrumental broadening (FWHM=85$\rm{km} \ \rm{s}^{-1}$).  Line flux
measurements were made by comparing the Gaussian fit to the
non-variable star S0-2 between 2.17-2.18 $\mu$m.  Statistical
uncertainties were taken from the root mean square (rms) error of
Gaussian fits to three independent data subsets; however, we note that
systematic errors (e.g. OH sky lines) are not included.

The final \brgamma line maps were used to evaluate G2's physical size
and absolute astrometric position.  The size of G2's \brgamma emission
was estimated by comparing the FWHM of G2 to that of nearby stars.  
To track the motions of G2 in the plane of
the sky, its OSIRIS position was transformed into an absolute
coordinate system.  To construct an absolute reference frame, we used the
precise locations of well characterized stars in the field 
(see figure \ref{fig:kn3}), as measured with \emph{StarFinder} 
\citep{Diolaiti-etal00}.  The position of G2 from the
two-dimensional Gaussian fit was added to the list of positions for
each epoch.  Then, these positions were matched to those from LGS AO
$K'$ observations of the same region at a nearby epoch using a 
second-order polynomial transformation to account for translation,
rotation, and pixel scale differences between the images\footnote{In
  2012, only a linear transformation was used because fewer reference
  stars were available due to the smaller field of view (7 versus 19-27
  in other epochs).}.  Once the positions in the OSIRIS frames were
transformed into a $K'$ epoch, they were transformed into an absolute
reference frame in which SgrA*-radio is at rest at the origin, as
originally described in \citet[][]{Yelda-etal10} and updated in Yelda
et al. \emph{in prep}.  We investigated possible systematic
astrometric effects in OSIRIS by cross-checking the OSIRIS astrometry
of S0-2 to our standard NIRC2 analysis. The results are consistent,
validating the use of OSIRIS astrometry.

\subsection{NIRC2 Imaging Measurements}\label{sec:lp}
\begin{figure}[ht!]
    \centering 
    \subfigure[A collapsed OSIRIS cube from May 2006 showing the continuum
      sources though the narrow-band Kn3 filter.  The position of
      G2, which has no detectable $K$ continuum, is indicated with an
      arrow.] { \includegraphics[width=0.45\textwidth]{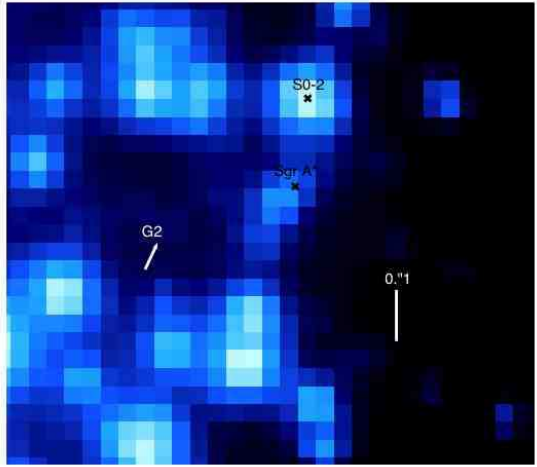}
      \label{fig:kn3}
    }
    \hspace{0.1in} 
    \subfigure[The Kn3 image shown in (a) is
      continuum-subtracted such that only \brgamma emission at G2's
      redshift is shown.  G2 is isolated in this continuum-subtracted
      map, minimizing the effect of stellar confusion on positional
      measurements.] {
      \includegraphics[width=0.45\textwidth]{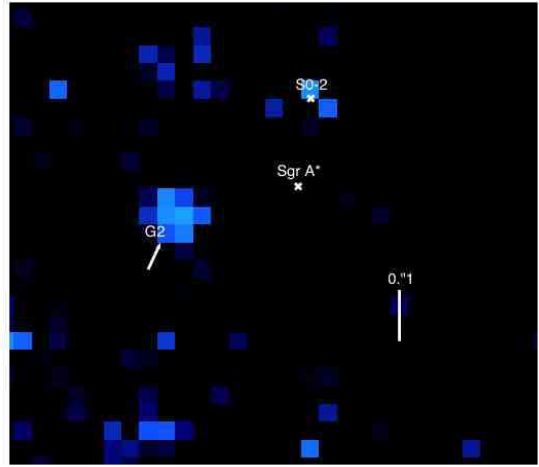}
        \label{fig:brgamma}
    } 
    \\ 
    \subfigure[A NIRC2 $K'$ image from May 2010 with the OSIRIS
      position of G2 indicated by a $3 \sigma$ contour.  For
      reference, the detections of other $K'$ sources are
      shown.  This image was used to derive G2's $K'$ magnitude
      upper limit since it is most isolated in this
      imaging epoch.]  {
      \includegraphics[width=0.45\textwidth]{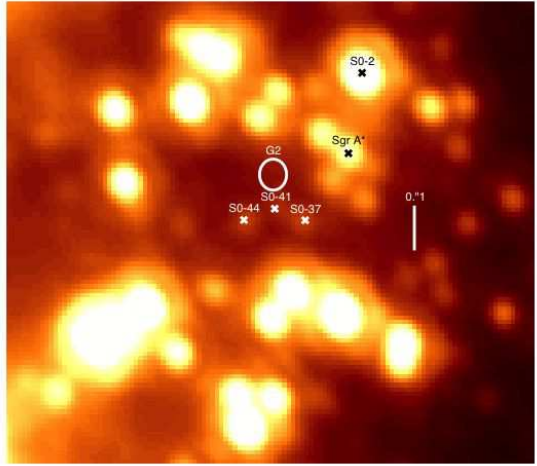}
      \label{fig:kp}
    }
    \hspace{0.1in}
    \subfigure[A NIRC2 $L'$ image from July 2012 which has been
      deconvolved and restored with a Gaussian Beam.  The background
      estimation from \emph{StarFinder} has also been added back in to
      show the dust filament which biases G2's $L'$ position in early
      epochs.  This feature extends upwards from the lower left of the
      image.  G2 is blended with another point source in 2012.] {
      \includegraphics[width=0.45\textwidth]{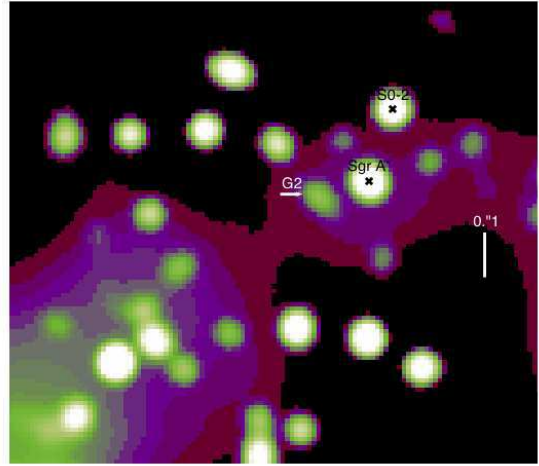}
      \label{fig:lp}
    }
    \caption{Images of the Galactic Center showing the \brgamma and
      $L'$ detections of G2, as well as the map in which the $K'$
      upper limit was derived.}
    \label{fig:images}
\end{figure}

The $L'$ position of G2 was estimated from calibrated
near-infrared images in a similar fashion to that described above for
the \brgamma line maps.  One difference was that we
deconvolved the individual images using the Lucy-Richardson algorithm
\citep{Richardson72,Lucy74,Lucy-Hook92} in order to help isolate the
point sources from the extended $L'$ dust emission.  Both the
large-scale background and the PSF were estimated using
\emph{StarFinder}.  After beam-restoring the individual frames with a
Gaussian having FWHM of half the theoretical resolution limit (40
mas), the frames were averaged to create a final image for each epoch
(see figure \ref{fig:lp}).  The results of this analysis do not vary
if the image is restored using a wider Gaussian beam.  To locate stars 
used as astrometric reference sources, \emph{StarFinder} was
run on the combined (deconvolved) frame using the beam-restoring
Gaussian as the reference PSF.  G2's position was determined from
an elliptical two-dimensional Gaussian fit\footnote{In 2012, a
  two-component elliptical Gaussian was fit since G2 is blended with
  another point source.}.  The fitting of an elliptical Gaussian
allows for extended structure and is analogous to the method used for
the \brgamma detection.  The positions for G2 from this method
are consistent with those produced via \emph{StarFinder}.  
The size of G2 was measured by comparing to the PSF of stars with a 
similar flux.  The
observed $L'$ positions were transformed into an absolute coordinate
system as described for the \brgamma detections.

One complication for measuring the $L'$ position of G2 in the plane of
the sky is that G2 was superimposed on a filament of dust emission
during the early- to mid- 2000's, which caused significant
astrometric bias (see figure \ref{fig:lp}).  To quantify this
effect, we performed a series of star-planting simulations using
a 2012 May $L'$ image, in which G2 is well off the filament.  In each
of the four simulations, an extra point source ($L' = 14$) was
planted at the expected G2 position at the time of earlier $L'$
observations (2003.440, 2004.567, 2005.580, and 2006.385) based upon
the orbit fit to the \brgamma measurements.  The background dust
emission does not have noticeable motions in the plane of the sky,
thus the astrometric bias at these locations in the 2012 image should
be representative of the bias at earlier epochs.  The position of
the artificial point source was extracted using \emph{StarFinder}.

To constrain the spectral energy distribution of the object, we
measured the integrated flux of G2 in the $L'$ frame
\citep[e.g.][]{Witzel-etal12}.  In contrast with \citet{Eckart-etal13},
but in agreement with \citet{Gillessen-etal12}, we detect no $K'$
counterpart for G2; therefore, we performed star planting simulations
to determine the $K'$ upper limit.  We used the \brgamma astrometry of G2 to
predict the positions of G2 in the $K'$ data.  
Then, an artificial point source was planted at G2's predicted
position in the May 2010 image, when G2 is most isolated (see figure
\ref{fig:kp}).
Using a modified version of the
\emph{StarFinder} algorithm (Boehle et al. \emph{in prep}), we
searched for a point source within a 3 pixel box centered on the
planted star, given the PSF originally extracted from the
May 2010 image.

\subsection{Orbital Modeling} \label{sec:orb}
We fit an orbit to G2's astrometric positions and RV measurements.
$L'$-astrometric and RV measurements reported by
\citet{Gillessen-etal12,Gillessen-etal13} were also included; however,
we fit the \brgamma and $L'$ astrometry separately.  We assumed G2
follows a purely Keplerian orbit around the central black hole.

The orbit inference problem aims to find the most likely orbital
parameters (eccentricity, period, time of periapse passage,
inclination, position angle of the ascending node, and the longitude
of periapse) given the data. In addition to G2's orbital parameters,
the parameters of the gravitational potential need to be inferred
(mass and distance to black hole, and its perceived 2D position and 3D
velocity). Since the observed motion of G2 does not entail enough
information to constrain all these parameters, the orbit of S0-2,
which has undergone a full orbit, was fit simultaneously and
effectively determines the gravitational potential
\citep[see][]{LMeyer-etal12}.
Compared to earlier publications, we
amended our Keplerian orbit fitting code to use the Bayesian sampler
MultiNest \citep{Feroz-Hobson08}.  We verified the results are
equivalent with our previous Monte Carlo approach
\citep[e.g.][]{Ghez-etal08}.

\section{Results} \label{sec:results}
\begin{figure*}[ht!]
\centering
\includegraphics[width=6in]{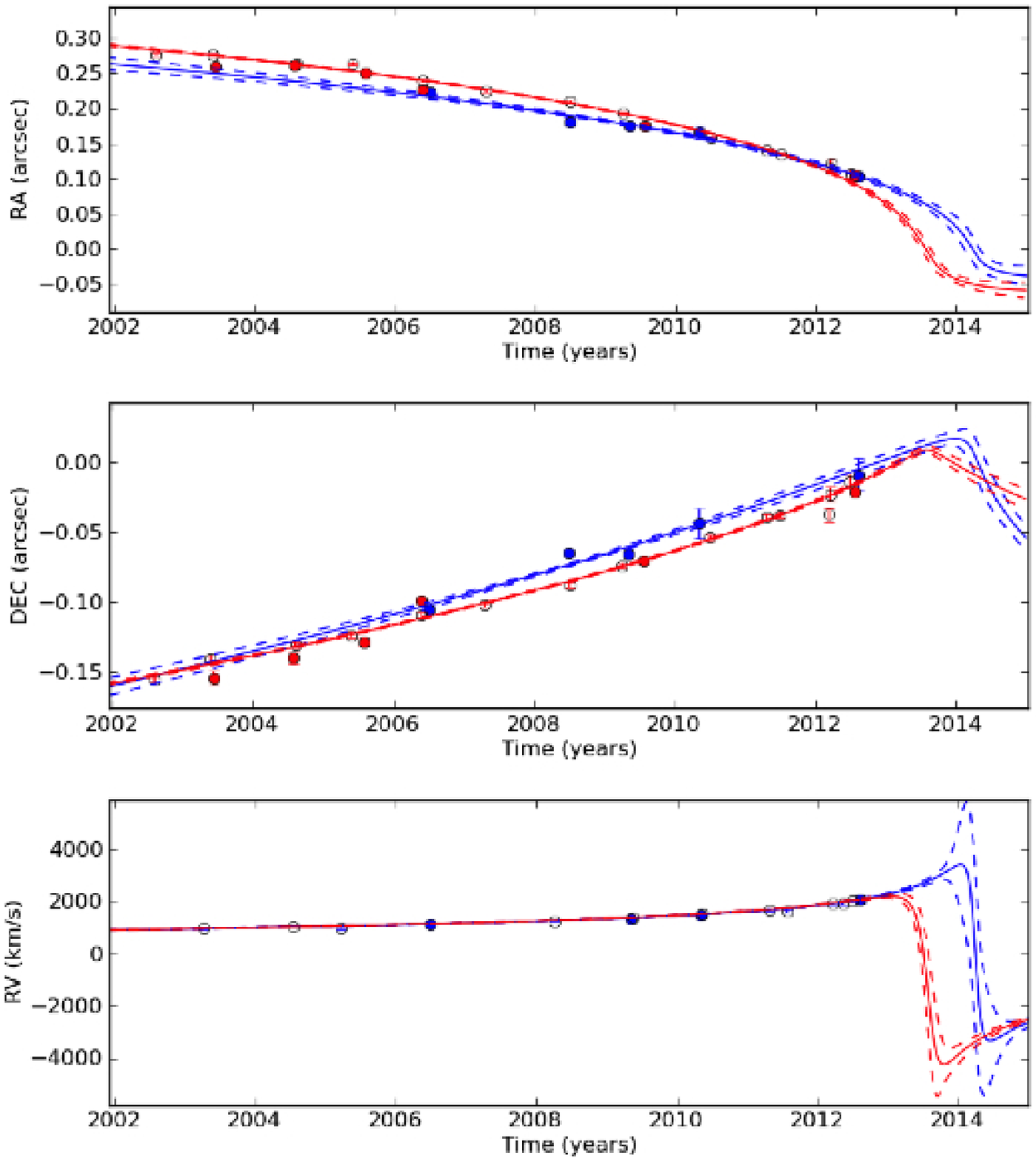}
\caption{The orbital fit using the \brgamma astrometry is shown in
  blue and the fit derived using the $L'$ astrometry is shown in red.
  Both solutions use radial velocity fits determined from the \brgamma
  emission line.  Astrometric points obtained from data presented in
  this paper are indicated with filled points, and open circles
  indicate measurements presented in
  \citet{Gillessen-etal12,Gillessen-etal13}.  Both $L'$ data sets
  appear offset from the \brgamma astrometry.}
\label{fig:orb}
\end{figure*}

Six key results emerge from our analysis of G2:
\begin{itemize}
\item \emph{Apparent positional offset between the \brgamma line
  emission and $L'$ continuum emission of G2:} The $L'$ astrometry is
  not consistent with our \brgamma astrometry, which is reported here
  for the first time (see figure \ref{fig:orb}).  The positional
  difference can be explained by astrometric bias in the $L'$
  astrometry.  In our star planting analysis, we find the extracted
  position of an artificial $L'$ source on the dust filament where G2
  is located at early epochs differs from the input positions by an
  average of $\sim$1.2 pix (12 mas).  This is almost an order of
  magnitude larger than typical centroiding uncertainties at $L'$ and
  the astrometric bias found in a region isolated from extended
  background flux.  We therefore exclude $L'$ astrometry from our
  primary orbital fits.
\item \emph{Revised Orbital Parameters:} The Keplerian orbit derived
  using \brgamma astrometry differs from the orbit derived using $L'$
  astrometry \citep[e.g.][]{Gillessen-etal13} by more than can be
  explained by formal measurement uncertainties (see figure
  \ref{fig:orb}).  We list the most likely orbital parameters from \brgamma 
  in table \ref{tab:orb}.  This new orbit for G2 has pushed the closest
  approach date to March 2014, has a closer periastron ($130
  \ \rm{AU}$, $1900 R_{\rm{s}}$) and a higher eccentricity ($0.9814
  \pm 0.0060$).
\item \emph{G2's compact size:} G2's observed size is comparable to
  the Keck angular resolution in almost all our observations; however,
  it exhibits marginal spatial extent in our highest quality
  spectroscopic data set.  At this time (2006), the observed
  half-width at half-maximum is $20 \pm 4 \ \rm{mas}$ after accounting
  for the observed PSF of S0-2 ($\rm{FWHM} = 74 \pm 3 \ \rm{mas}$).
  G2 also shows a velocity gradient along its orbit, with the most
  highly redshifted portion of the object closest to the black hole,
  consistent with what is expected for a tidal interaction event.  The
  observed spatial extent is consistent with that presented in
  \citet{Gillessen-etal12}, thus the \brgamma spatial extent of G2
  along the direction of motion is relatively constant and corresponds
  to a size on the order of $\sim$100 AU.
\item \emph{Spectral Evolution:} We find an increasing \brgamma line
  width for G2, with FWHM measurements ranging from $135 \pm 30
  \ \rm{km} \ \rm{s}^{-1}$ in 2006 to $706 \pm 250 \ \rm{km}
  \ \rm{s}^{-1}$ in 2012 (see table \ref{tab:obs}).  These
  measurements are consistent with those presented by
  \citet{Gillessen-etal12,Gillessen-etal13}.
\item \emph{Constant Brightness:} Both at $L'$ and \brgamma the flux
  of G2 is constant within uncertainties.  The integrated
  flux of G2 at $L'$ has an average value of $14.9 \pm 0.3 \,
  \rm{mag}$.  The \brgamma line fluxes in 2006, 2009, 2010, and 2012
  are $0.884 \pm 0.116\%$, $0.919 \pm 0.098\%$, $1.13 \pm 0.41\%$,
  $1.14 \pm 0.87\%$ of S0-2's continuum flux between 2.17-2.18
  \micron, respectively.
\item \emph{Source Color:} Our star planting simulations indicate that
  the upper $K'$ magnitude limit of G2 is $K'=20$ and is $\sim 2$
  magnitudes deeper than the earlier upper $K'$ limit imposed on G2
  \citep{Gillessen-etal12}.  The deeper $K'$ limit yields a dust
  temperature below $\sim$500K if G2 is a pure gas cloud.  If G2 has
  an underlying stellar source, a $K'$ magnitude limit places distinct
  limits on the luminosity of the star.  This limit is similar to the
  expected $K'$ magnitude for a low-mass T Tauri star in the Galactic
  Center \citep[e.g.][]{Scoville-Burkert13}, and eliminates a more
  massive star unless it is shrouded by dust which self-extincts the
  star at $K'$.
\end{itemize}

\begin{deluxetable}{lc}
\tabletypesize{\scriptsize}
\tablecaption{Orbital elements for G2\tablenotemark{a}\label{tab:orb}}
\tablewidth{0pt}
\tablehead{\colhead{Parameter} & \colhead{Value\tablenotemark{b}}}
\startdata
Time of closest approach ($T_0$)                 & $2014.21 \pm 0.14 \ \rm{yrs}$ \\
Eccentricity ($e$)                               & $0.9814 \pm 0.0060$ \\
Period ($P$)                                     & $276 \pm 111 \ \rm{yrs}$\\
Angle to periapse ($\omega$)                     & $88 \pm 6 \ \rm{deg}$ \\
Inclination ($i$)                                & $121 \pm 3 \ \rm{deg}$ \\
Position angle of the ascending node ($\Omega$)  & $56 \pm 11 \ \rm{deg}$ \\
\enddata
\tablenotetext{a}{Parameters describing the gravitational potential are found in \citet{LMeyer-etal12}.}
\tablenotetext{b}{Values provided are the mean and standard deviation of the marginalized one-dimensional distributions.}
\end{deluxetable}

\section{Discussion} \label{sec:disc}
The two orbits derived by \citet{Gillessen-etal12,Gillessen-etal13}
differ from each other by more than can be explained with formal
measurement uncertainties.  Systematic errors in the L' astrometry
naturally explain the discrepancy.  Because the offset becomes less
significant and the RV measurements play an increasingly dominant role
in the orbital modeling, the measurements seem to converge on the
solution obtained with the less biased \brgamma astrometry and RV
measurements.

There are several implications of the new orbital model parameters.
First, the revised periapse date of March 2014 and the closer
periastron are important for the design and interpretation of SgrA*
monitoring programs designed to test for an increased accretion flow.
With a special NRAO call for proposals and tremendous attention called
to this object, which was originally anticipated to experience
periapse in June 2013, more than 30 programs have been approved for
the summer of 2013 covering radio to $\gamma$-ray wavelengths.  While
some models predict that a cloud would generate enhanced radio
emission from SgrA* well in advance of periapse passage
\citep[e.g.][]{Narayan-etal12,Sadowski-etal13b,Sadowski-etal13a}, none
has yet been detected \citep{Kassim-etal13}, although this may not yet
be expected with the revised periapse passage.

Second, the geometry of the new orbit may call two previously
suggested associations into question.  Most importantly, the previous
claimed ``tail" of low surface brightness emission, which was central
to the original claim that G2 is a pure gas cloud, is no longer
securely affiliated with the compact ``head", which has been analyzed
in this work.  The new orbit also falls 3.5$\sigma$ off the plane of
the disk of young stars orbiting the central black hole at $\sim$0.05
- 2 pc
(\citealt{Genzel-etal00,Levin-Beloboradov03,Paumard-etal06,Lu-etal09,Bartko-etal09};
Yelda et al. \emph{in prep}) which has been suggested as the possible
origin of the gas cloud through colliding stellar winds of young stars
\citep{Ozernoy-etal97,Cuadra-etal06} or a central stellar source
\citep[e.g.][]{Murray-Clay-Loeb12}.

Third, the even higher eccentricity provides strong constraints on
G2's origin.  It is notable that a high eccentricity is consistent
with the outcome of three body exchanges between a binary system and
the central black hole which may explain the dense concentration of B
stars (the S-stars) in the innermost regions of the Galaxy
\citep[e.g.][]{Alexander-Livio04}.  If a triple star system is invoked
rather than a binary, G2 could possibly be the result of a recent
merger between two components to produce the observed physical
properties (red, compact, and at times marginally resolved) and an
exchange with the central black hole with the remaining component to
produce the observed orbital parameters (highly eccentric, short
period).  While we favor a stellar model, this question will be soon
addressed, as G2 should remain intact through its periapse if this is
indeed correct.

\acknowledgments The authors would like to acknowledge the invaluable
feedback from conference attendees at workshops where this work has
been presented (AAS 221st Meeting, Keck 20th Anniversary Celebration,
Galactic Nuclei Ringberg Workshop).  Support for this work was
provided by NSF grant AST-0909218.  Data presented herein were taken
at the W. M. Keck Observatory. The W. M. Keck Observatory is operated
as a scientific partnership among the California Institute of
Technology, the University of California, and the National Aeronautics
and Space Administration. The Observatory was made possible by the
generous financial support of the W. M. Keck Foundation.

\emph{Facilities}: Keck: II (LGS AO, NIRC2, OSIRIS), Keck: I (LGS AO,
OSIRIS)

\end{document}